\documentclass[]{aastex631}

\shorttitle{GALE - MHD Modeling}
\shortauthors{Stejko et al.}
\graphicspath{{./}{figures/}}

\begin{document}

\title{Constraining Global Solar Models through Helioseismic Analysis}

\author[0000-0001-7483-3257]{Andrey M. Stejko}
\affiliation{New Jersey Institute of Technology, Newark, NJ 07012, USA}
 
\author[0000-0003-0364-4883]{Alexander G. Kosovichev}
\affiliation{New Jersey Institute of Technology, Newark, NJ 07012, USA}

\author[0000-0002-2256-5884]{Nicholas A. Featherstone}
\affiliation{SWRI, Department of Space Studies, Boulder,
CO 80302, USA}

\author[0000-0002-2671-8796]{Gustavo Guerrero}
\affiliation{New Jersey Institute of Technology, Newark, NJ 07012, USA}
\affiliation{Universidade Federal de Minas Gerais, Belo Horizonte, MG, 31270-901, Brazil}

\author[0000-0001-7612-6628]{Bradley W. Hindman}
\affiliation{JILA, University of Colorado, Boulder, CO 80309-0440, USA}
\affiliation{Applied Mathematics, University of Colorado, Boulder, CO 80309-0526, USA}

\author[0000-0001-9001-6118]{Loren I. Matilsky}
\affiliation{JILA, University of Colorado, Boulder, CO 80309-0440, USA}
\affiliation{Applied Mathematics, University of Colorado, Boulder, CO 80309-0526, USA}

\author[0000-0002-9292-4600 ]{J\"orn Warnecke}
\affiliation{Max-Planck-Institut für Sonnensystemforschung, Justus-von-Liebig-Weg 3, D-37077 G\"ottingen, Germany}





\begin{abstract}

\indent Global hydrodynamic simulations of internal solar dynamics have focused on replicating the conditions for solar-like (equator rotating faster than the poles) differential rotation and meridional circulation using the results of helioseismic inversions as a constraint. Inferences of meridional circulation, however, have provided controversial results showing the possibility of one, two, or multiple cells along the radius. To resolve this controversy and develop a more robust understanding of global flow regimes in the solar interior, we apply a ``forward-modeling'' approach to the analysis of helioseismic signatures of meridional circulation profiles obtained from numerical simulations. We employ the global acoustic modeling code GALE to simulate the propagation of acoustic waves through regimes of mean mass flows generated by global hydrodynamic and magnetohydrodynamic models: EULAG, the Pencil Code, and the Rayleigh code. These models are used to create synthetic dopplergram data products, used as inputs for local time-distance helioseismology techniques. Helioseismic travel-time signals from solutions obtained through global numerical simulations are compared directly with inferences from solar observations, in order to set additional constraints on global model parameters in a direct way. We show that even though these models are able to replicate solar-like differential rotation, the resulting rotationally-constrained convection develops a multi-cell global meridional circulation profile that is measurably inconsistent with local time-distance inferences of solar observations. However, we find that the development of rotationally-unconstrained convection close to the model surface is able to maintain solar-like differential rotation, while having a significant impact on the helioseismic travel-time signal, replicating solar observations within one standard deviation of the error due to noise.

\end{abstract}



\section{Introduction}\label{sec:intro}

The implementation of non-linear hydrodynamic (HD) and magnetohydrodynamic (MHD) modeling is often contrasted with mean-field simulations, which have found success in replicating solar processes and building out models of the global dynamics that drive solar mean mass flows \citep{1989drsc.book.....R,1995A&A...299..446K,2004ARep...48..153K,2018ApJ...854...67P} and the generation of the global solar dynamo---see \citet{2010LRSP....7....3C} for a comprehensive review. These models, however, often require ad-hoc prescriptions of internal solar parameters, resulting in potentially unrealistic distributions and amplitudes of turbulent transport coefficients, along with as of yet unknown mechanisms that may have significant impacts on mean mass flows. Non-linear HD/MHD modeling attempts to replicate global solar flows through a more holistic development of global dynamics and the solar dynamo, by simulating convective energy transport in simplified models of solar plasma. Non-linear global modeling, in particular, has made tremendous strides since the seminal works of \citet{1972SoPh...27....3G} and \citet{1981ApJS...46..211G}. This can be seen most clearly in simulations analyzing the conditions for solar-like differential rotation \citep[e.g.,][]{2013ApJ...779..176G,2014MNRAS.438L..76G,2014ApJ...789...35F,2015ApJ...804...67F,2019ApJ...871..217M,2020ApJ...898..111M,2020A&A...642A..66W,2022arXiv220204183H}.\\
\indent In-depth investigations have been made possible due in part to the success of global helioseismology in mapping the rotational structure of the solar interior \citep[e.g.,][]{1997SoPh..170...43K,1998ApJ...505..390S,2002ApJ...567.1234S,2011JPhCS.271a2061H} providing detailed constraints for solar models to replicate. Reliably inferring the Sun's internal meridional circulation, however, has remained a difficult problem. Local time-distance helioseismology techniques have had significantly more trouble probing into deeper parts of the solar convection zone ($r<0.96R_{\odot}$). Large-scale systematic errors such as the center-to-limb (CToL) effect \citep[see][]{2012ApJ...749L...5Z,2019_chen}, and apparent downflows in magnetic regions \citep{2015ApJ...805..165L}, have proven challenging to disentangle effectively---resulting in widely varying conjectures on the structure of meridional circulation in the solar convection zone (SCZ). This has culminated in a disagreement over whether meridional circulation exhibits a single-cell \citep{2020Sci...368.1469G} or a double/multi-cell structure \citep{2013ApJ...774L..29Z,2014ApJ...784..145K,2019_chen}. Progress has steadily been made, however, with the development of new approaches to disentangling the CToL effect using frequency-dependent analysis \citep{2019_chen, 2020ASSP...57..107R}.\\
\indent Recent helioseismic analysis of synthetic dopplergram data generated by global acoustic models have shown that the noise in time-distance measurements is too high to make pronouncements on whether meridional circulation has more than one cell \citep{2021ApJ...911...90S}. Helioseismic observations, however, can still be useful in setting constraints on global, non-linear, convectively-driven models in a limited capacity. Even though differentiating between single-cell and multi-cell structures remains difficult, we can gauge how well the particular multi-cell structure commonly exhibited by MHD/HD models in solar-like rotational regimes
\citep[e.g.,][]{2013ApJ...779..176G,2013ApJ...778...41K,2015ApJ...804...67F, 2019ApJ...871..217M,2020ApJ...898..120H,2020A&A...642A..66W} agrees with solar observations, and what critical insights can be gained from global models of turbulent solar convection. We apply a ``forward-modeling'' method to compare helioseismic travel-time signatures of these models directly to solar observations \citep{2021ApJS..253....9S,2021ApJ...911...90S}---computing travel-time differences using local time-distance helioseismic techniques on synthetic dopplergram data. This data is created using a global acoustic code that computes oscillations over background velocities imported from non-linear convectively-driven models. Comparing the resulting travel-time differences to those taken from observational full-disk dopplergram data results in a more direct comparison of measured signals without relying on inversions to estimate velocity profiles.\\
\indent This paper is organized as follows. In \S\ref{sec:model} we briefly describe the computational set-up of the acoustic simulation code and time-distance helioseismic analysis procedure, used to generate and analyze synthetic dopplergrams. In \S\ref{sec:MM} we present results of helioseismic measurements for convectively-driven Models R1x, M5, \& H38 (described therein), and in \S\ref{sec:MC}, we analyze the helioseismic signatures generated by models with a varying stratification (N3 \& N5). \S\ref{sec:comparison} presents a comparison of results with solar observations, and finally, in \S\ref{sec:conclusions} we offer an analysis and discussion of how these results can be employed as constraints on the future development of global convectively-driven solar models.

\section{Acoustic Modeling and Helioseismic Analysis} \label{sec:model}

\indent A compressible 3D acoustic simulation code (GALE; \citealt{2021ApJS..253....9S}) is used to generate synthetic dopplergrams for the forward-modeling analysis of convectively-driven hydrodynamic global models. This algorithm employs novel pseudo-spectral methods to offer an efficient and flexible platform for computing the contributions of 3D background flow structures to acoustic perturbations within the simulated solar interior. The Euler equations are solved in their conservative form, in a fully spherical domain: $0 < \theta < \pi$, $0 < \phi \le 2\pi$, $0 < r \le 1.001R_{\odot}$.

\begin{equation}\label{eq:gov1}
  \frac{\partial \rho'}{\partial t} + \Upsilon' = 0 \ ,
\end{equation}
\begin{equation}\label{eq:gov2}
  \frac{\partial\Upsilon'}{\partial t} + \mathcal{M}' = -\nabla^{2}\left(p'\right) - \nabla\cdot\left(\rho'\tilde{g}_{r}\mathbf{\hat{r}}\right) + \nabla\cdot S\mathbf{\hat{r}} \ ,
\end{equation}
\begin{equation}\label{eq:gov3}
  \frac{\partial p'}{\partial t} = - \frac{\Gamma_{1}\tilde{p}}{\tilde{\rho}}\left(\mathcal{O}_{p} + \tilde{\rho} \mathbf{u}'\cdot\frac{N^{2}}{g}\mathbf{\hat{r}}\right) \ .
\end{equation}

\indent Linear perturbations in the potential field are computed, with solenoidal contributions discarded. This is achieved with a split-field formulation, by computing the divergence of the momentum field ($\Upsilon = \nabla\cdot \rho \mathbf{u}$). The governing equations are then linearized by solving for perturbations (denoted by a prime) from base parameters (denoted by a tilde) of pressure ($p$), density ($\rho$), gravity ($g$), the Brunt-Väisälä frequency ($N^{2}$), and the adiabatic ratio ($\Gamma_{1}$). Contributions of the divergence of the material derivative are denoted by $\mathcal{M}'$, and adiabatic contributions to the conservation of energy by $\mathcal{O}_{p}$. 3D solar oscillation data is generated for user-specified background flow profiles ($\tilde{\mathbf{u}}$), reproducing shifts to the solar oscillation spectrum \citep{2021ApJS..253....9S}. This algorithm implements hybrid MPI and OpenMP protocols that enable massively parallelized computation. The pseudo-spectral computational method works by decomposing field terms into vector spherical harmonics (VSH) and tensor spherical harmonics (TSH), allowing for the efficient computation of symmetric second-order tensor and dyad terms. Realization noise, mimicking the kind seen in observational data, is simulated through a stochastic excitation of source terms ($S$) generated by a chi-squared distribution of random frequency perturbations in the top $0.1\%$ of the solar interior.\\
\indent We create synthetic dopplergrams using global 3D background velocity profiles generated by the non-linear convection simulations of the solar interior performed with the EULAG code, the Rayleigh code, and the Pencil Code (see Section \ref{sec:MM} for more detailed descriptions of each code and complementing dataset), allowing us to characterize the influence of each distinct regime on helioseismic measurements. The GALE code is initialized to a maximum spectral resolution of $\ell_{\rm{max}} = 200$---high enough to sample the convective interior up to $r \sim 0.96R_{\odot}$. The acoustic wave-field is evolved for approximately 67 hours of model time, generating synthetic dopplergram data sampled from the Model approximately $300\text{ km}$ above the solar surface ($R_{\odot} =6.9599\times10^{10}\text{ cm}$). Such a time-scale is too short to effectively resolve the travel-time signal from realization noise \citep{2008ApJ...689L.161B}, so we leverage the dependence of the signal-to-noise ratio (S/R) on the square root of the temporal sampling window, increasing background velocities by a factor of 25 \citep{2013ApJ...762..132H}, effectively simulating approximately 5 years of observations.\\
\indent To analyze the resulting synthetic dopplergram data, we employ the local time-distance helioseismology technique described by \citet{2009ApJ...702.1150Z} and \citet{2021ApJ...911...90S}. This method allows global mean flows in the solar interior to be inferred by measuring their impact on the acoustic wave-field. Waves traveling in opposite directions along p-mode ray paths will exhibit travel-time differences when moving through a medium with some average velocity along their path. These travel-time differences are calculated from a cross-correlation of two points on the solar surface, through the process of Gabor wavelet fitting \citep{1997ASSL..225..241K}, where a wave-packet function \citep[see][]{2021ApJS..253....9S} is fit to the measured signal using the iterative Levenberg-Marquardt method. Each pixel in the synthetic dopplergram, between a latitude range of $50^{\circ}$N and $50^{\circ}$S, is treated as the center of a $60^{\circ}\times 60^{\circ}$ patch that is remapped into an azimuthal equidistant projection (Postel's projection) at a resolution of approximately $0.6^{\circ}$ per pixel. The radius of concentric circles drawn on this patch represents half of the angular distance along the model surface ($\Delta$) that the acoustic ray travels, penetrating a maximum depth ($r_{i}$) at the center of the patch that can be estimated as $r_{i} = c(r_{i})L/\omega$, where $c$ is the sound speed, $\omega$ is the angular frequency and $L = \sqrt{l(l+1)}$, is effectively the spherical harmonic degree. Pixels are selected along these circles at every interval ($1.2^{\circ}$) for $12^{\circ}$ to $42^{\circ}$, corresponding to an approximate depth of $r_{i} = 0.93 R_{\odot} - 0.72 R_{\odot}$. $30^{\circ}$-wide sectors in the North and South (1 pixel in radius) are then averaged and cross-correlated with each other in order to compute the travel-time differences created by mean meridional flows in the global model. These travel-time differences can be used to infer internal solar velocities using the ray-path approximation \citep{1999_giles} or Born approximation kernels that estimate the depth dependence of velocity contributions \citep{2001ESASP.464..187B}. In this analysis, we compare travel-time measurements obtained from global convection simulations directly to those computed from solar observations (Section \ref{sec:MC}), without the need to rely on approximations made using inversion techniques.

\section{Analyzing Meridional Profiles of Convectively Driven Models}\label{sec:MM}

\begin{figure}[htb]
\center
\includegraphics[width=15cm]{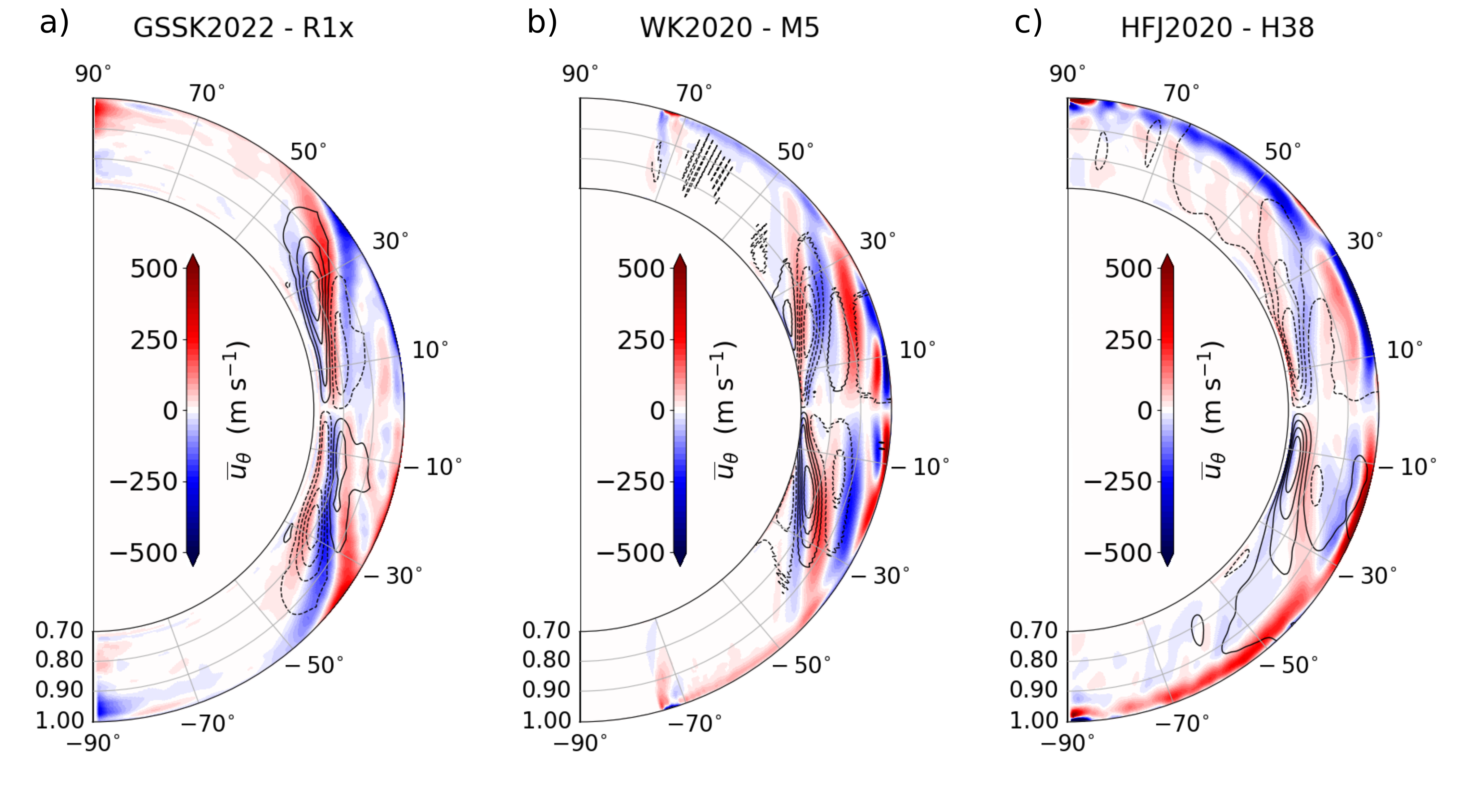}
\caption{Latitudinal velocities ($\tilde{u}_{\theta}$) amplified to a maximum of $500\text{ m/s}$. Meridional circulation profiles are shown corresponding to a solar-like differential rotation generated by convectively driven models of EULAG (a), the Pencil Code (b, \citealt{2020A&A...642A..66W}), and Rayleigh (c, \citealt{2020ApJ...898..120H}). Solid and dashed contours represent counterclockwise and clockwise circulation, respectively.}
\label{fig:MC_SL}
\end{figure}

\indent We compare three meridional profiles generated by the non-linear convectively driven HD/MHD codes: EULAG \citep{2013JCoPh.236..608S}, the Pencil Code \citep{2012ApJ...755L..22K, 2018A&A...616A..72W, 2020A&A...642A..66W,PencilCode}, and the Rayleigh Code \citep{2016ApJ...818...32F,featherstone_nicholas_a_2021_5774039}. The first meridional circulation profile that we analyze is described as Model R1x by \citet{guerrero_22}, generated using the hydrodynamic (without the magnetic field) global Model EULAG, where the anelastic approximation is used to simulate convection in a global computational domain measuring: $0 \le \phi \le 2\pi$, $0 \le \theta \le \pi$, $0.60 R_{\odot} \le r \le 0.964 R_{\odot}$. In this model, convection is primarily driven by a super-adiabatic state, prescribed with an ambient potential temperature function for an ideal gas, whose polytropic index corresponds to marginally unstable convection ($m<1.5$). The index is chosen to mimic the density stratification prescribed by the Solar Model S \citep{1996Sci...272.1286C} within the computational domain---corresponding to approximately $N_{\rho} = 3.64$ density scale heights in the simulated convection zone ($r = 0.70 R_{\odot} - 0.964 R_{\odot}$). The angular rotation rate of the reference frame is set to slightly above the average solar rotation rate ($\Omega_{0} = 1.17 \Omega_{\odot} $), sufficient to maintain solar-like differential rotation, at a global Rossby number of $\text{Ro }= 0.56$ in the convection zone (with a radial extent of $H = 0.964 R_{\odot} - 0.72 R_{\odot}$), calculated as $\text{Ro } = (2\Omega_{0}\tau_{c})^{-1} $, where $\tau_{c} = H/u_{\rm{rms}}$, is the convective turnover time. The resulting meridional circulation profile can be seen in panel (a) of Fig. \ref{fig:MC_SL}.

\indent The Pencil Code \citep{PencilCode} is a high-order finite-difference algorithm used for the computation of compressible magnetohydrodynamics on highly parallelized computational architectures. The full code and instructions for its use and installation are maintained on https://github.com/pencil-code/pencil-code. This code has been employed for global dynamo simulations in a wedge geometry represented by a quarter-spherical mesh grid measuring: $0 \le \phi \le \pi/2$, $\pi/12 < \theta < 11\pi/12$, $0.70R_{\odot} < r \le R_{\odot}$. The computational set-up is described in detail by \citet{2013ApJ...778...41K}. This algorithm is used to simulate the development of the global solar dynamo by simulating heat flux (${\partial T}/{\partial r}$) at the bottom boundary and prescribing a radiative heat conductivity profile that falls off with increased radius ($K \sim r^{-15}$). This model has been used to investigate the rotational dependence of global solar properties \citep[e.g.][]{2018A&A...616A..72W,2020A&A...642A..66W}, evincing solar-like differential rotation at higher rotation rates---represented by a Rossby number of $\text{Ro} < 0.27$, where $\text{Ro} = (2\Omega_{0}\tau_{c})^{-1}$ and $\tau_{c}$ is defined as the convective turnover time, averaged over the entire computational domain. We analyze Model M5 \citep{2020A&A...642A..66W} with an angular rotation rate of $\Omega = 5\Omega_{\odot}$ and a Rossby number of $\text{Ro} < 0.12$. This model is actuated with a large-scale magnetic field, influencing the development of meridional flow. The resulting meridional circulation profile can be seen in panel (b) of Fig. \ref{fig:MC_SL}.

\begin{figure}[htb!]
\center
\includegraphics[width=15cm]{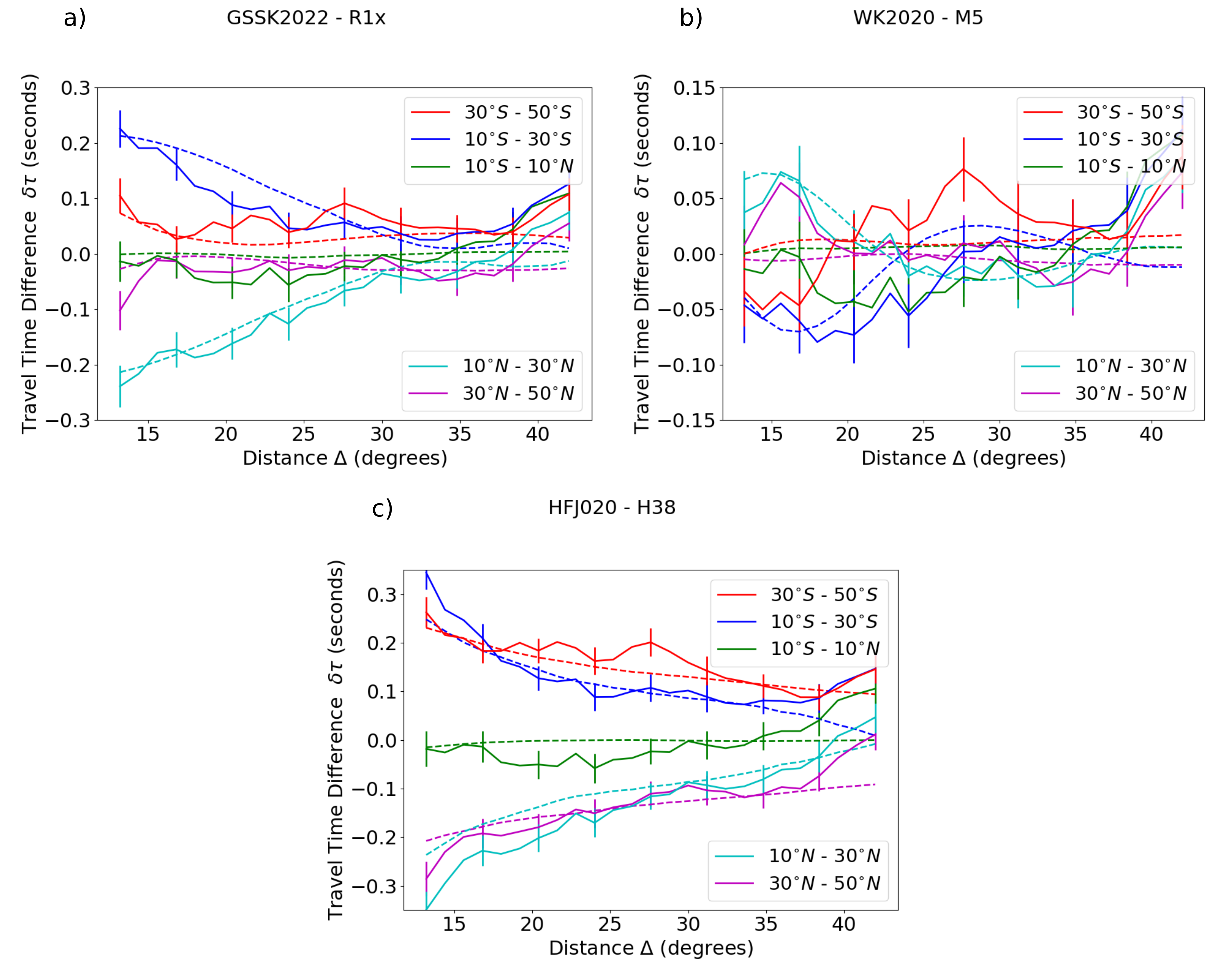}
\caption{The N-S travel-time differences ($\delta\tau_{NS}$) as a function of travel distance ($\Delta$) for Models R1x (a), M5 (b), H38 (c). The travel-time measurements are phase-speed filtered ($\sigma = 0.05v_{p}$; \citealt{2007ApJ...659.1736N}) for 5 latitudinal averages. Dashed lines are theoretical times computed using the ray-path approximation \citep{1999_giles}. Error bars show the standard deviation of the measured noise ($\sigma_{NS}$, Equation \ref{eq:error}).}
\label{fig:TD_SL_F}
\end{figure}

\indent Rayleigh is a highly parallelized pseudo-spectral algorithm used to simulate convection in stellar interiors under the anelastic approximation \citep[see][for details]{2016ApJ...818...32F}. The full code, as well as its instructions for use and installation, are made freely available at https://github.com/geodynamics/Rayleigh and through \citet{featherstone_nicholas_a_2021_5774039}. Background states are represented by the adiabatic stratifcation of an ideal gas with a polytropic index of $n= 1.5$, defined as a function of density scale heights throughout the domain ($N_{\rho}$). The number of scale heights can be freely altered to simulate various stratification regimes, with $N_{\rho} = 3$ most closely resembling the density profile in the solar Model S \citep{1996Sci...272.1286C} inside the simulated radial range ($0.718R_{\odot} \le r \le 0.946R_{\odot}$). Convection is driven by a constant deposition of energy into the domain, with a linear radial dependence on internal pressure. The dispersal of this energy occurs via heat conduction at the upper boundary, set by the stellar luminosity parameter $L_{*}$. Models are computed on a fully spherical shell ($0 \le \phi \le 2\pi$, $0 < \theta < \pi$, $0.718R_{\odot} \le r \le 0.946R_{\odot}$). The meridional circulation profile for Model H38 \citep{2020ApJ...898..120H} can be seen in panel (c) of Fig. \ref{fig:MC_SL}. This model is characterized by three density scale heights ($N_{\rho} = 3$), a bulk flux Rayleigh number of $\text{Ra} = 8.61\times10^{5}$ \citep[see][for details]{2020ApJ...898..120H}, and a Rossby number of $\text{Ro} = 5.94\times10^{-2}$---exhibiting a solar-like differential rotation at the edge of antisolar (equator rotating slower than poles) transition, as well as a multi-cell arrangement of its meridional circulation profile. The Rossby number is calculated as $\text{Ro} = u_{\rm{rms}}(2\Omega_{0}H)^{-1}$, where $u_{rm{rms}}$ is the RMS velocity integrated over the full spherical shell, with radial size $H$.

\indent Models R1x and H38 are stretched in order to match the radial extent of the solar surface ($r = R_{\odot}$), with the size of each radial mesh point multiplied by a constant value. This slightly changes the background stratification of the models, moving them further from inferred solar stratification profiles \citep{1996Sci...272.1286C}. This is not a significant concern, however, as these models don't purport to faithfully replicate turbulent convective parameters on the Sun, and the resulting global flow profiles are still unable to completely reproduce solar dynamics. Stretching these profiles, however, allows for a better interpretation of the helioseismic signals that these simulated regimes would generate if they reached the surface---giving us a better idea of the constraints that can be placed on such models with time-distance helioseismology. Meridional velocities ($u_{r}, u_{\theta}$) are amplified to a maximum of $500\text{ m/s}$---a 25-fold increase surface velocities that peak at of maximum velocity of $\sim 20\text{ m/s}$ and average out to $\sim 10 - 17\text{ m/s}$ in regions of interest on the Model surface \citep[see][]{2018A&A...611A..92R}.

\begin{figure}[htb!]
\center
\includegraphics[width=15cm]{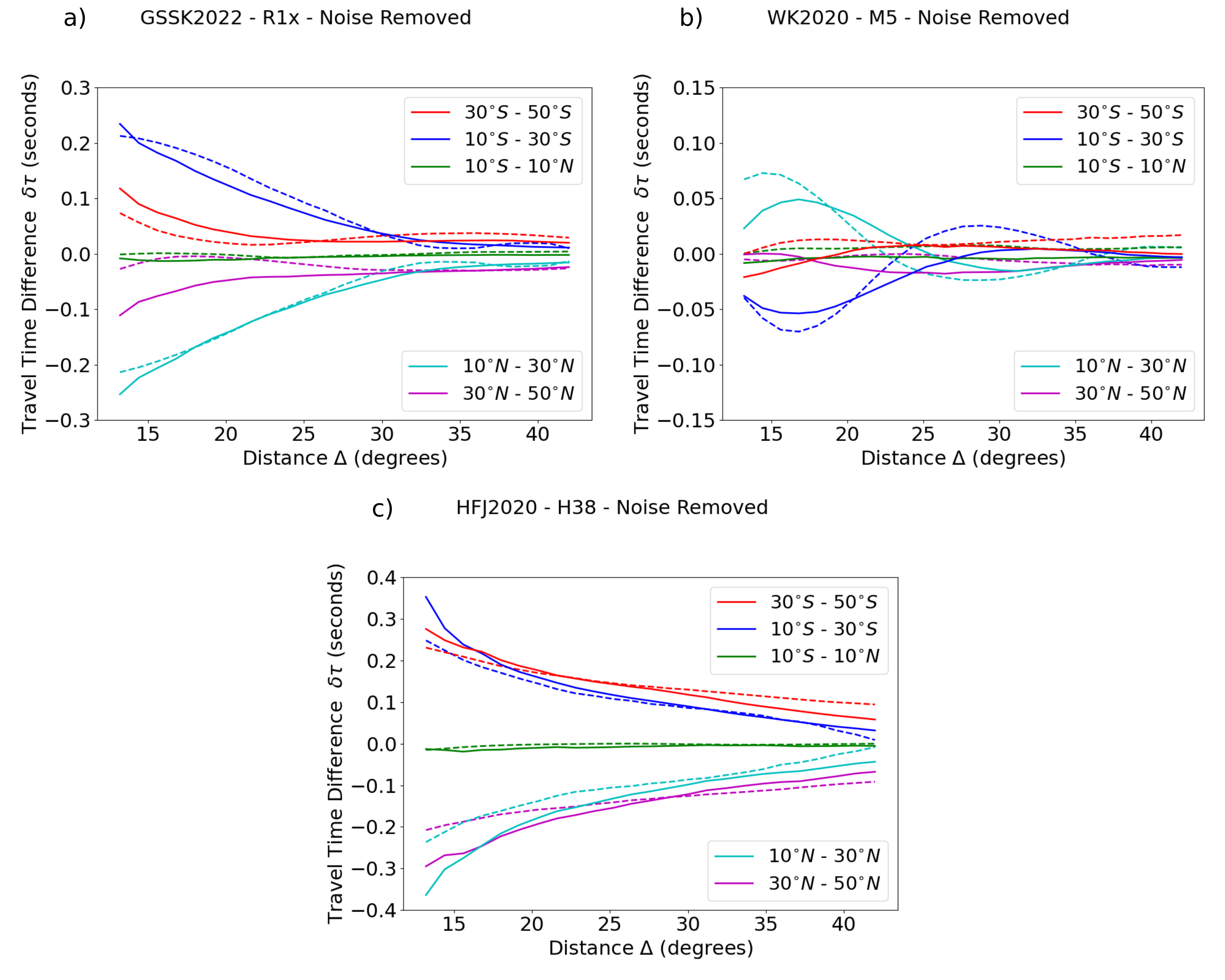}
\caption{The N-S travel-time differences ($\delta\tau_{NS}$) as a function of travel distance ($\Delta$) for Models R1x (a), M5 (b), H38 (c). The travel-time measurements are phase-speed filtered ($\sigma = 0.05v_{p}$; \citealt{2007ApJ...659.1736N}), with profiles of noise subtracted for 5 latitudinal averages. Dashed lines are theoretical times computed using the ray-path approximation \citep{1999_giles}.}
\label{fig:TD_SL_NR}
\end{figure}

\indent The meridional circulation profiles are characterized by Taylor columns strongly aligned with the rotational axis, indicative of the models' inability to break the Taylor-Proudman balance \citep{1989drsc.book.....R,1995A&A...299..446K,2005ApJ...622.1320R} in fast-rotating regimes. These low-latitude columnar cells correspond to cylindrical convective modes, known as banana cells or Busse columns \citep{Busse1970}, that are seen to develop in convectively-driven solar and stellar simulations \citep[e.g.,][]{2011A&A...531A.162K,2013ApJ...779..176G,2015ApJ...804...67F, 2020ApJ...898..120H}. A prominent feature of such models is shown in their inability to develop strong continuous circulation cells that stretch across the latitudinal extent such as those inferred in solar observations of sub-surface meridional circulation \citep{2013ApJ...774L..29Z,2013ApJ...778L..38S,2014ApJ...784..145K,2018ApJ...860...48L,2020Sci...368.1469G}. This is true of Models R1x and M5, as well as other solar-like differential rotation models generated by the Rayleigh code \citep[see][]{2020ApJ...898..120H}. Model H38 appears to deviate from this trend, however, showing mostly continuous poleward flows with a small exception around the $45^{\circ}$ latitude---a potential result of the limited rotational constraint (especially near the surface) on the model. Another feature commonly observed in these models is the formation of multiple cells with return flows very close to the model surface, with many managing to break through to the upper boundary. This has a very noticeable effect on the helioseismic signature which can be visualized by plotting N-S travel-time differences ($\delta\tau_{NS}$) as a function of their travel distance ($\Delta =  12^{\circ} - 42^{\circ}$)---corresponding to turning points: $r = 0.93R_{\odot} - 0.72R_{\odot}$, respectively, in the solar interior. In order to reduce noise in our measurements, we plot latitudinal averages of these travel-time differences in places that approximately correspond with similar continuous features seen in the models, expressed by the five following ranges: $30^{\circ}\text{N} - 50^{\circ}\text{N}$, $10^{\circ}\text{N} - 30^{\circ}\text{N}$, $10^{\circ}\text{S} - 10^{\circ}\text{N}$, $10^{\circ}\text{S} - 30^{\circ}\text{S}$, $30^{\circ}\text{S} - 50^{\circ}\text{S}$. The travel-time differences for the three regimes of meridional circulation (R1x, M5, H38) can be seen in Fig. \ref{fig:TD_SL_F}, where travel-time differences sampled from our synthetic dopplergram data are shown as solid lines, and are compared with the expected travel-time differences computed using the ray-path approximation (see dashed lines, \citealt{1999_giles}). The model travel-time differences are divided by the same factor of 25 that velocity values ($u_{r}$, $u_{\theta}$) were amplified by---justified by their close match to the linear ray-path approximation (seen in Fig. \ref{fig:TD_SL_NR}). Error-bars represent one standard deviation of the measured noise, computed by sampling 100 synthetic dopplergrams generated with a unique source function on a model with no background flows \citep{2021ApJ...911...90S}. This noise can be removed in order to more clearly visualize the travel-time differences (shown in Fig. \ref{fig:TD_SL_NR}), by subtracting travel-times computed from a synthetic dopplergram initiated with the same source function ($S$, Equation \ref{eq:gov2}), but without any background flows.

\indent It is immediately evident that the helioseismic response is quite a bit weaker at all latitudinal averages when compared to travel-time differences measured from solar observations (see Fig. \ref{fig:TD_MDI} in Section \ref{sec:comparison}). The signal for model M5 (panel b, Fig. \ref{fig:TD_SL_NR}) consistently hovers near zero at all depth ranges, with the slightest increase in strength at the $10^{\circ}-30^{\circ}$ range, coinciding with the development of strong rotational-axis aligned meridional circulation \citep{2020A&A...642A..66W}. These flows are unable to form cohesive latitudinal cells, with the polar surface flows constantly switching directions as strong columnar convective cells penetrate from the interior. It becomes apparent that the quick succession of flow reversals associated with such multi-cell structures negates average helioseismic responses very rapidly with depth. It becomes difficult to distinguish an extremely weak global meridional flow, from a very strong one with near-surface reversals, as we move deeper into the convection zone. Another peculiar feature we observe is an oscillatory pattern of the signal with depth, seen clearly in model M5 and very slightly in model R1x. It appears that one of the marks of a strong columnar multi-cell arrangement of the meridional flow can be rapid increases/decreases in travel-time differences that move between positive and negative values for the same latitudinal range. These reversals are smaller than the level of noise seen in observations, making them difficult to detect, however, they demonstrate a potential signature of a strong, rotationally constrained, multi-cell meridional flow. Model R1x shows a similar signal, however, with a stronger response in the $10^{\circ}-30^{\circ}$ latitudinal range. While the columnar convection deeper in the model interior appears similar to Model M5, the poleward meridional flow on the solar surface is more continuous, possibly due to the lower rotational velocity of the model. This results in a measurably stronger signal in this range. This effect is even more pronounced in Model H38 where a continuous surface latitudinal cell is allowed to form in the $10^{\circ}-30^{\circ}$ range, however, it is difficult to compare these two signals directly as the scaling of the average surface flow strength is very different in the two models ($\sim 10\text{ m/s}$ for model R1x, and $\sim 17\text{ m/s}$ for model H38). Even though the surface flow cell is stronger and more cohesive in Model H38, the travel-time signal is not significantly stronger than in model R1x, as a near-surface return flow quickly negates its impact as we probe deeper into the interior. Model H38 is also characterized by a significant difference at higher latitudes, allowing strong (mostly) poleward flows to form throughout the entire extent of the model surface. This feature appears to be unique to the minimally rotationally constrained model, even among the other simulations of \citet{2020ApJ...898..120H}.\\
\indent These models demonstrate that the strongest helioseismic responses are seen in regions where large continuous poleward flows are allowed to form on the surface and penetrate deeper into the model interior. While the arrangement of internal convective cells do show unique signals, their helioseismic responses may be indistinguishable within the noise of lower half of the SCZ ($r < 0.85R_{\odot}$). Helioseismic constraints may not be able to tell us exactly how many circulation cells there are or their specific arrangement. These results, however, demonstrate that a strong baseline for surface flow speeds, combined with a drop-off in travel-time differences with depth, results in a good indication of whether a strong cohesive polar flow extends deep into the solar interior, as well as how likely a possible near-surface return flow is---as seen in the global helioseismic analysis of \citet{2007AN....328.1009M} and the correlation tracking done by \citet{2012ApJ...760...84H}. We explore this question in greater detail in the next two sections by analyzing the effect of a changing return flow height on the helioseismic signal.

\section{Analysis of Models with Varying Stratification Regimes}\label{sec:MC}

\begin{figure}[htb]
\center
\includegraphics[width=10cm]{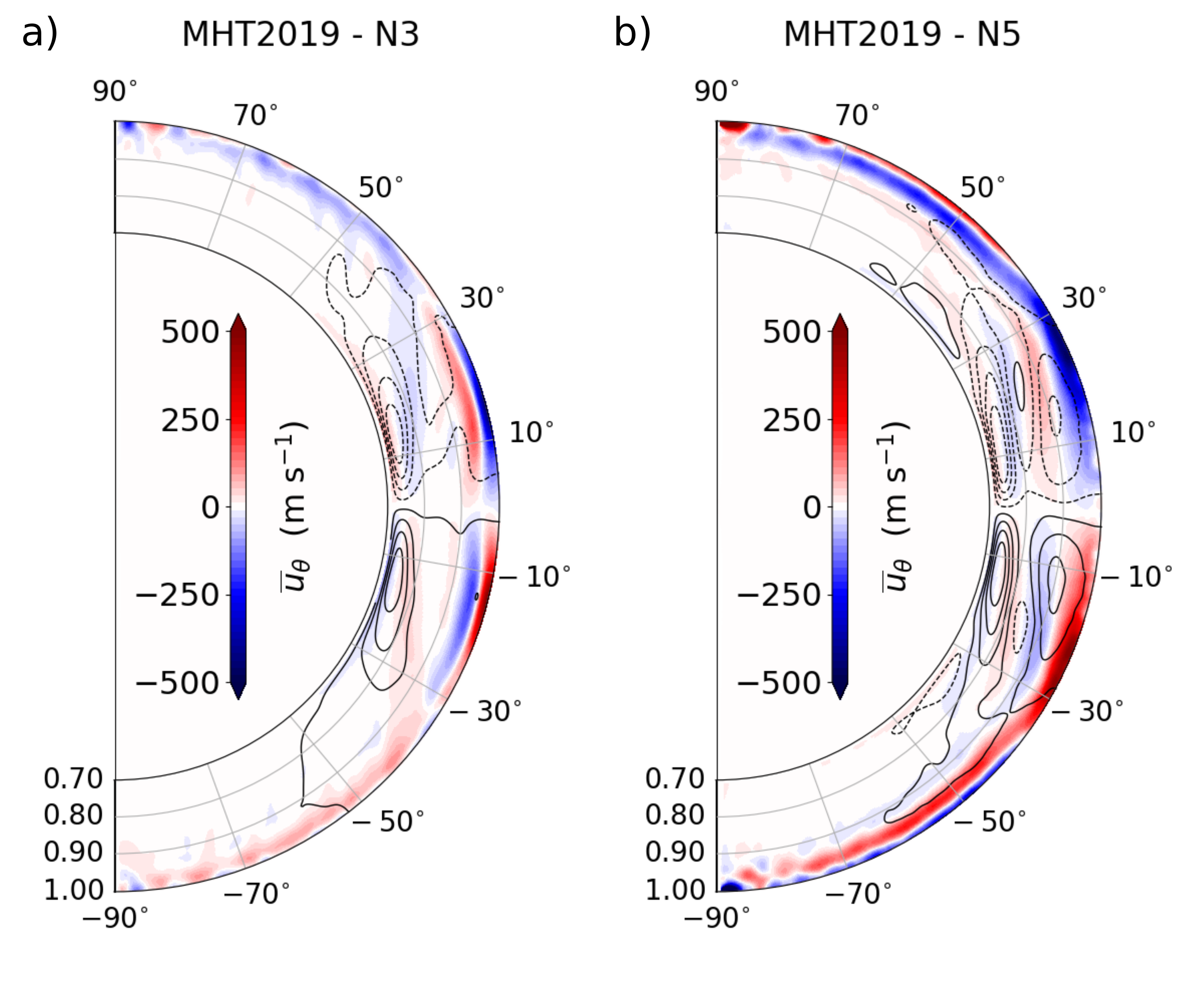}
\caption{Latitudinal velocities ($\tilde{u}_{\theta}$) amplified to a maximum of $500\text{ m/s}$. Meridional circulation profiles are generated by the Rayleigh code \citep{2019ApJ...871..217M}, contrasting models with three density scale heights (a, N3) and five density scale heights (b, N5). Solid and dashed contours represent counterclockwise and clockwise circulation, respectively.}
\label{fig:MC_n}
\end{figure}

\indent The effect of model stratification on the observed helioseismic signature is illustrated by comparing the two models of \citet{2019ApJ...871..217M} (N3 \& N5), generated by the Rayleigh code. The meridional circulation profiles of these models can be seen in Fig. \ref{fig:MC_n}, where, as in the previous section, the profiles are stretched to the solar surface ($R_{\odot}$). The models are actuated with identical input parameters, with the exception of the number of density scale heights; $N_{\rho} = 3$ for Model N3 and $N_{\rho} = 5$ for Model N5, resulting in bulk Rossby numbers of $\text{Ro} = 0.1345$ and $\text{Ro} = 0.4793$, respectively. An analysis of the local Rossby numbers (defined as $\text{Ro} = v'_{rms}(r)(2\Omega_{0}H_{\rho}(r))^{-1}$, where $v'_{rms}(r)$ is a spherically averaged RMS velocity and $H_{\rho}(r)$ is the local density scale height) by \citet{2019ApJ...871..217M}, shows that the increased near-surface stratification in Model N5 results in a rotationally unconstrained layer above $r/r_{o} \sim 0.97$, where $r_{o} = 6.586\times 10^{10}\text{ cm}$, in their model, and corresponds to the solar radius ($r_{o} = R_{\odot}$) after we stretch it. Angular momentum transport in this region is dominated by inwardly directed turbulent Reynolds stresses resulting from the increased convective transport of downflow plumes. The impact of angular momentum transport due to columnar convection (Busse columns) is limited to rotationally constrained convection in the interior, allowing for the development of a larger more continuous global meridional flow cell near the surface of Model N5.

\begin{figure}[htb!]
\center
\includegraphics[width=15cm]{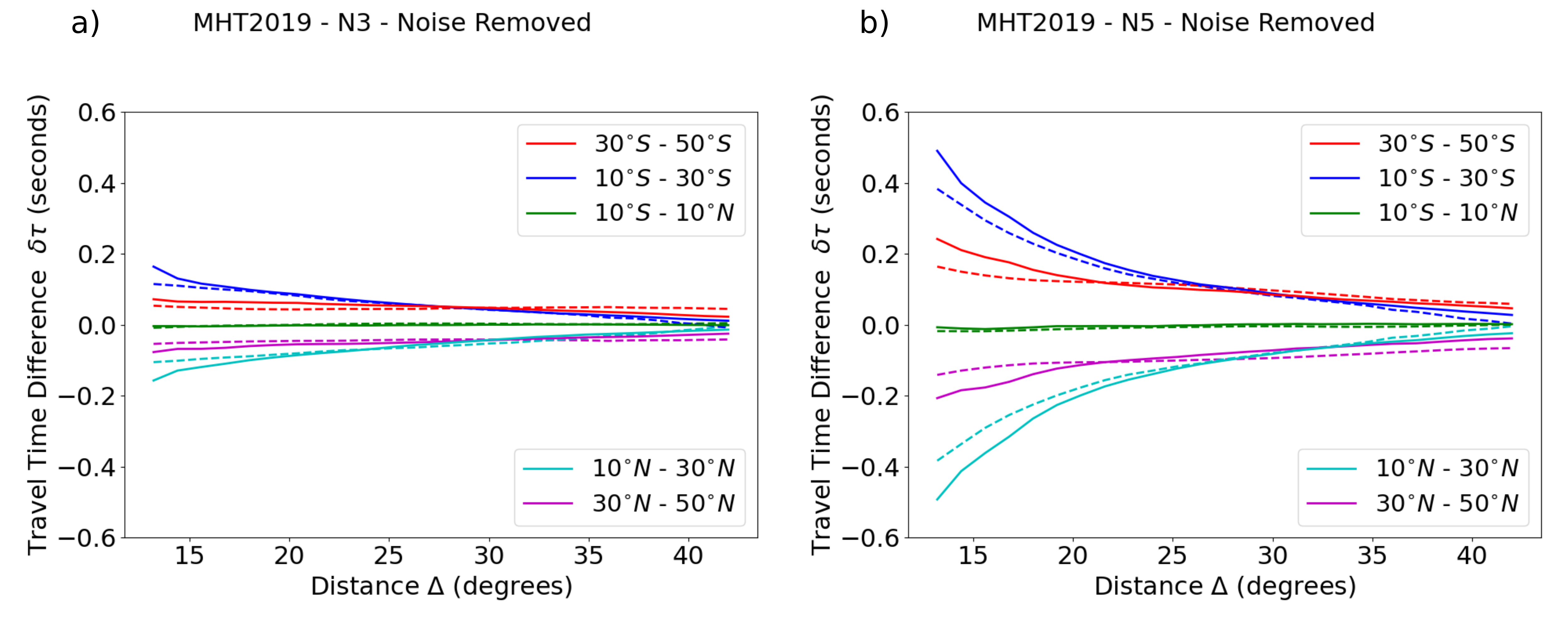}
\caption{The N-S travel-time differences ($\delta\tau_{NS}$) as a function of travel distance ($\Delta$) for Models N3 (a), N5 (b). The travel-time measurements are shown under a phase-speed filter ($\sigma = 0.05v_{p}$; \citealt{2007ApJ...659.1736N}), with profiles of noise subtracted for 5 latitudinal averages. Dashed lines are theoretical times computed using the ray-path approximation \citep{1999_giles}.}
\label{fig:TD_N_NR}
\end{figure}

\indent We analyze the resulting helioseismic signatures by plotting the same latitudinal travel-time differences as in Section \ref{sec:MM}---corresponding to the approximate extent of continuous latitudinal features in the profiles. Fig. \ref{fig:TD_N_NR} shows these latitudinal averages with the profile of noise removed for clarity. Solid lines correspond to the measured signal, while dashed lines are computed using the ray-path approximation \citep{1999_giles}. The enlarged primary surface meridional flow cell in Model N5 shows a significantly increased helioseismic response as compared to Model N3, as well as the other solar convection models (R1x, M5, H38). This is most pronounced at the $10^{\circ} - 30^{\circ}$ latitude that shows almost a three-fold increase in average travel-time differences ($\delta\tau_{NS}$) near the model surface. Higher latitudes  ($>|30^{\circ}|$) show a significantly diminished response due to the development of a reverse flow near the upper boundary of Model N5. This reversal corresponds to a meridional torque attempting to balance the inward transport of angular momentum due to the Reynolds stresses in the region \citep{2019ApJ...871..217M}. The resulting equator-ward surface flow at higher latitudes deviates from solar observations, requiring an as of yet unknown mechanism to balance it, and showing the need for a greater understanding of solar meridional flow structure. 

\indent Focusing on the $10^{\circ} - 30^{\circ}$ latitudinal range, shows that the presence of a large continuous poleward motion penetrating deeper into the model interior results in the most impactful change to the helioseismic signal, far outweighing the importance of any arrangement within the deep convective interior ($r < 0.90R_{\odot}$), whether that be the columnar multi-cell formations seen in Models R1x/M5, or the less rotational-axis aligned multi-cell meridional circulation of Models H38/N3. This inference is supported by results of previous works---showing that a double-cell meridional circulation profile, induced by a reversal near the base of the SCZ, shows slight differences when compared to a single-cell profile generated by the same mean-field solar model \citep{2021ApJ...911...90S}. A positive implication of these results is that, even though the internal arrangement of global flows may be inaccessible, results of local time-distance helioseismology can be effectively used to put constraints on the height of the initial flow reversal. We attempt to do this in the subsequent section by comparing the results of our helioseimic analysis with solar observations.

\section{Comparison with Helioseismic Observations}\label{sec:comparison}

\begin{figure}[htb]
\center
\includegraphics[width=8cm]{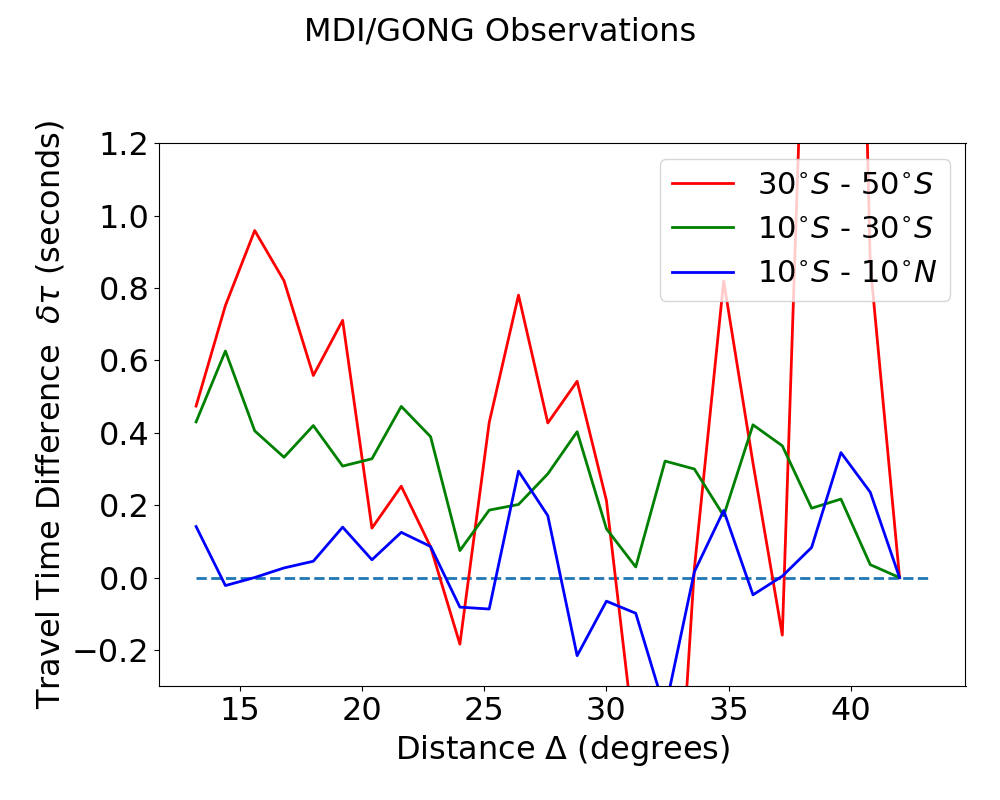}
\caption{The N-S travel-time differences ($\delta\tau_{NS}$) as a function of travel distance ($\Delta$) for MDI/GONG data published by \citet{2020Sci...368.1469G}. Three latitude ranges are shown: $10^{\circ}\text{S} - 10^{\circ}\text{N}$, $10^{\circ}\text{S} - 30^{\circ}\text{S}$, $30^{\circ}\text{S} - 50^{\circ}\text{S}$, averaged with their antisymmetric counterparts in the northern hemisphere in order to reduce noise.}
\label{fig:TD_MDI}
\end{figure}

In order to compare the results of our analysis directly to solar observations, we use the latest data from approximately 23 years of combined observations made by the Michelson Doppler Imager \citep[MDI,][]{1995SoPh..162..129S} of the Solar and Heliospheric Observatory (SOHO), as well as the Global Oscillation Network Group \citep[GONG,][]{1996Sci...272.1284H}. The observations are described in detail and published by \citet{2020Sci...368.1469G}; publicly available at the Open Research Data Repository of the Max Planck Society. We show travel-time differences (Fig. \ref{fig:TD_MDI}) for the same latitudinal averages as in our previous analyses. In order to reduce noise, we average the signals in both hemispheres, only showing $10^{\circ}\text{S} - 10^{\circ}\text{N}$, $10^{\circ}\text{S} - 30^{\circ}\text{S}$, $30^{\circ}\text{S} - 50^{\circ}\text{S}$.

\begin{figure}[htb]
\center
\includegraphics[width=15cm]{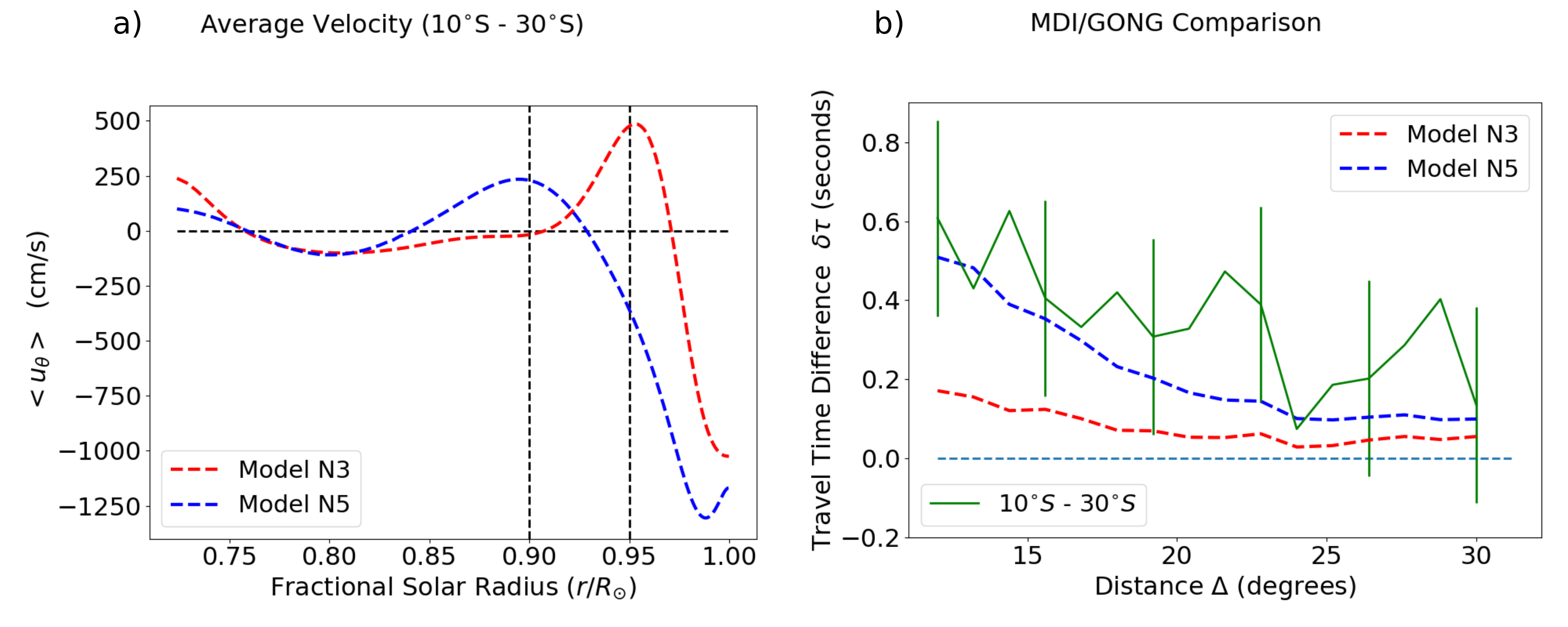}
\caption{a) The average latitudinal velocity $\left< u_{\theta}\right >$ in the $10^{\circ}\text{S} - 30^{\circ}\text{S}$ latitude range for Models N3 (red) and N5 (blue), stretched to the solar surface ($R_{\odot}$). b) The N-S travel-time differences ($\delta\tau_{NS}$) as a function of travel distance ($\Delta$) for MDI/GONG data published by \citet{2020Sci...368.1469G}. Latitude ranges in both hemispheres are averaged in order to reduce noise and are compared to dashed lines representing latitudinal averages for Models N3 (red) and N5 (blue). Error-bars are computed as the standard deviation ($\sigma_{MG}$, Equation \ref{eq:error}) of the travel-time differences in the $10^{\circ}\text{N} - 10^{\circ}\text{S}$ latitude range from zero.}
\label{fig:TD_MDI_c}
\end{figure}

\indent The noise at higher latitude ranges ($30^{\circ}\text{S} - 50^{\circ}\text{S}$) is too high to make significant pronouncements---especially at greater depths ($\Delta > 20^{\circ}$, $r<0.87R_{\odot}$), however, the near-surface regions do show a signal more consistent with a continuous circulation cell (Model H38) as opposed to the significant weakening/reversals seen in most of the other convectively-driven models (R1x, M5, N3, N5, Fig. \ref{fig:TD_SL_NR}, Fig. \ref{fig:TD_N_NR}). We concentrate further analysis on the $10^{\circ}\text{S} - 30^{\circ}\text{S}$ latitudinal range where noise is significantly reduced. The average latitudinal velocity in this range ($\left< u_{\theta}\right >$) is shown in panel (a) of Fig. \ref{fig:TD_MDI_c} for Models N3 and N5. The deep return flow cell structure (N5) begins to diverge from the shallow one (N3) at approximately $r = 0.80R_{\odot}$, showing the structure of two potential return flow profiles that culminate in an average surface velocity of approximately 10 m/s, with a maximum of 20 m/s. In panel (b) of Fig. \ref{fig:TD_MDI_c} we show the travel-time differences computed from synthetic dopplergrams (dashed lines) for Models N3 and N5 (see Fig. \ref{fig:TD_N_NR}) in this region, comparing them to MDI/GONG observations (\citealt{2020Sci...368.1469G}, solid line) in the same latitudinal range. The error-bars are computed as one standard deviation (Equation \ref{eq:error}) of the travel-time differences in the $10^{\circ}\text{N} - 10^{\circ}\text{S}$ latitude range from zero.

\begin{equation}\label{eq:error}
  \sigma_{MG} = \sqrt{\frac{1}{N}\sum_{i=1}^{N}  \delta\tau_{i}^{2}}\ .
\end{equation}

\indent The travel-time differences calculated for Model N5 are a better match for the observational signal at depths $r > 0.80R_{\odot}$. The average z-score (number of standard deviations from the mean) of the signal in this depth range is $z_{N5} =  0.522$ for Model N5, and $z_{N3} =  1.099$ for Model N3. Assuming that the travel-time difference measurements of Models N3 and N5 are the mean of a normal noise distribution with a standard deviation of $\sigma_{MG}$, the probability of measuring a signal at least as extreme (p-value) as the travel-time differences computed by \citet{2020Sci...368.1469G} ($10^{\circ}\text{S} - 30^{\circ}\text{S}$, $12 < \Delta < 30^{\circ}$) would be approximately $p_{N3} = 0.14$ for Model N3 and $p_{N5} = 0.30$ for Model N5---representing a significantly increased likelihood that the average depth of the maximum return flow in this latitude range is approximately at or slightly below $r \sim 0.90R_{\odot}$, assuming an accurate scaling of surface flows (10 m/s average with a 20 m/s maximum, see \citet{2018A&A...611A..92R}). Model R1x has a similar surface flow scaling with a corresponding p-value of $p_{R1x} = 0.17$. Models M5 and H38 are, however, difficult to compare directly as model M5 has a very small average surface velocity in this region ($\sim$ 5 m/s) and model H38 has a very large one ($\sim$ 17 m/s).\\
\indent The major differentiating factor in the models that we can compare, is in the depth of the primary circulation cell on the model surface. Whether the arrangement is a strong columnar multi-cell structure, such as in Model R1x, or a more latitudinal one with weaker columnar cells, as in Model N3, they exhibit minimal travel-time differences---showing a very small drop-off with depth, with the addition of realization noise leaving the signals appearing nearly horizontal (see Fig. \ref{fig:TD_SL_F}). A comparison with solar observations shows that this is unlikely to be the case, as measured travel time-differences show a very distinct slope, larger than the error associated with the measurement (Fig. \ref{fig:TD_MDI}). We show that such a slope is most strongly associated with the radial extent of the primary circulation cell, corresponding to a minimal depth of the maximum return flow at $\sim 0.90 R_{\odot}$. This result, however, is strongly dependent on the proper estimate of meridional flow strength on the solar surface, which is variable, requires long temporal sampling windows to accurately gauge \citep{1996ApJ...460.1027H,2010ApJ...725..658U,2016LNP...914...25K,2018A&A...611A..92R}, and becomes unreliable at higher latitudes ($\pm 45^{\circ}$) owing to projection effects such as foreshortening.

\section{Discussion}\label{sec:conclusions}

\indent Global convectively-driven hydrodynamic models are unlikely to be able to recreate the conditions of the solar interior any time soon. The extreme dynamics of turbulent solar convection---with dimensional parameter estimates of the Reynolds numbers in the range of: $\text{Re} \sim 10^{10}-10^{13}$ and a Rayleigh number of $\text{Ra} \sim 10^{20}$ \citep[see][]{2006JFM...563...43R}, preclude a full understanding of the organization of buoyant injection and turbulent dissipation in the large range of turbulent scales, leaving us with a necessity of estimating the action of sub-grid scale turbulent dissipation with techniques such as the Large Eddy Simulation (LES) implementation of the dynamic Smagorinsky model \citep{germano91}. Other techniques to simulate the dissipation rate and the cascade of energy in the inertial subrange use implicit methods (e.g., ``implicit'' large eddy simulation or ILES, \citealt{iles07}), which simulates turbulent dissipation via a truncation of high-order terms in the computational scheme. While such methods have found success in replicating the organization of turbulence in direct numerical simulations (DNS) \citep{elliott02}, it is difficult to gauge if they are a realistic proxy for the solar interior. This is especially true considering the high stratification of solar plasma and the large range of energetic scales. This becomes a greater concern near the solar surface, along with an increasing velocity and compression and radiation effects that are too computationally expensive to model. This can be problematic for numerical simulations, as it is becoming increasingly apparent that these upper layers may be necessary in order to fully replicate global solar dynamics \citep{2020ApJ...888...16S}.\\
\indent Even though global modeling has its limitations, it remains a useful tool for understanding the action of chaotic systems within defined parameters. Rather than trying to simulate exact solar conditions, parameters can be tweaked to create a more robust understanding of the set of conditions that result in dynamical behaviors observed on the Sun. In hydrodynamic regimes this means reproducing global mean flow patterns such as differential rotation and meridional circulation. To that end we demonstrate how forward-modeling the helioseismic inferences of meridional circulation can be used as an additional constraint on global MHD/HD models, and improve our understanding of the convective turbulent parameters and stratification profiles needed to more accurately simulate solar conditions. We show that the multi-cell arrangement commonly associated convection simulations that reproduce solar-like differential rotation \citep[e.g.,][]{2013ApJ...779..176G,2013ApJ...778...41K,2014ApJ...789...35F,2020A&A...642A..66W,2022arXiv220204183H}, exhibits a weak travel-time difference signal with a strong curvature that can oscillate around zero with increasing depth. This signal seems to be characteristic of the development of rotationally-constrained columnar convection at low latitudes. A more linear drop-off curve is seen in models that develop more latitudinal flows near the surface, with weaker internal columnar convection (H38 \& N3), however, they still show a small helioseismic response. Increasing the radial extent of the primary circulation cell at the model surface causes the most pronounced impact, allowing convective models (N5) to most closely replicate solar observations. This impact is greater than one standard deviation of the realization noise, making it an effective constraint on mean meridonal flows generated by MHD/HD models. While these models are far from accurate simulations of solar dynamics, understanding the character of their global mean-mass flows can point us in the direction of more realistic solar simulations. Extending 3D global simulations to the solar surface is currently computationally unfeasible, however, this analysis reinforces the idea that the increased density stratification in near-surface layers may be necessary to adequately replicate global solar processes.

\begin{acknowledgments}
AMS would like to thank the heliophysics modeling and simulation team at NASA Ames Research Center for their support. This work is supported by the NASA grants: 80NSSC19K0630, 80NSSC19K1436, NNX14AB70G, NNX17AE76A, 80NSSC20K0602, 80NSSC17K0008, 80NSSC18K1125, 80NSSC19K0267, 80NSSC20K0193, 80NSSC19K1428, NNX13AG18G, NNX16AC92G, NNX17AG22G, 80NSSC18K1127. JW acknowledges funding by the European Research Council
(ERC) under the European Union's Horizon 2020 research and innovation programme (grant agreement n:o 818665 ``UniSDyn''). The authors thank the NASA Drive Science Center studying the Consequences Of Fields and Flows in the Interior and Exterior of the Sun (COFFIES) for providing multidisciplinary collaboration opportunities in heliophysics.
\end{acknowledgments}

%

\vspace{5mm}
\facilities{NASA AMES (NAS HECC)}


\software{GALE \citep{2021ApJ...911...90S},  
          EULAG-MHD \citep{2013JCoPh.236..608S}, 
          Rayleigh \citep[https://github.com/geodynamics/Rayleigh;][]{featherstone_nicholas_a_2021_5774039,2016ApJ...818...32F},
          The Pencil Code \citep[https://github.com/pencil-code/pencil-code;][]{PencilCode}
          }
          




\bibliography{Codes}{}
\bibliographystyle{aasjournal}



\end{document}